\documentstyle[twocolumn,aps,prl,epsf]{revtex}  
\draft
\title{\boldmath Observation of two time scales in the ferromagnetic manganite La$_{1-x}$Ca$_x$MnO$_3$, $x \approx 0.3$}
\author{R. H. Heffner,$^1$ J. E. Sonier,$^1$ D. E. MacLaughlin,$^2$ G. J. Nieuwenhuys,$^3$ G. Ehlers,$^4$ \\ F. Mezei,$^1$ S-W. Cheong,$^5$ J. S. Gardner$^1$ and H. R\"oder$^1$}
\address{$^1$Los Alamos National Laboratory, Los Alamos, New Mexico 87545}
\address{$^2$Department of Physics, University of California, Riverside, California 92521}
\address{$^3$Kamerlingh Onnes Laboratory, Leiden University, P. O. Box 9504, 2300 RA Leiden, The Netherlands}
\address{$^4$Institute Laue Langevin, 38042 Grenoble Cedex 9, France}
\address{$^5$Department of Physics and Astronomy, Rutgers University, Piscataway, NJ 08854}

\author{\small(version date \today)}			
\address{\parbox{14cm}{\bigskip\rm\small			
We report new zero-field muon spin relaxation and neutron spin echo measurements in ferromagnetic (FM) (La,Ca)MnO$_3$ which taken together suggest two spatially separated regions in close proximity possessing very different Mn-ion spin dynamics. One region corresponds to an extended cluster which displays 'critical slowing down' near $T_C$ and an increasing volume fraction below $T_C$. The second region possesses more slowly fluctuating spins and a decreasing volume fraction below $T_C$. These data are discussed in terms of the growth of small polarons into overlapping regions of correlated spins below $T_C$, resulting in a microscopically inhomogeneous FM transition.
\\[6pt] PACS numbers: 75.30.Vn, 75.40.Gb, 76.75.+i., 25.40.Fq}}

\begin{document} \maketitle		
\thispagestyle{myheadings}
\markright{\hfill {\small
LA-UR-99-4122\quad p.}\hspace{1mm}}

After several years of study it is clear that the richness in the temperature-composition phase diagram for La$_{1-x}$Ca$_x$MnO$_3$ \cite{Cheong1} is produced by a strong interplay between the spin, charge and lattice degrees of freedom in these materials \cite{Tokura}. Of particular interest has been the ferromagnetic (FM) insulator-to-metal transition and its accompanying large magnetoresistance. Millis and collaborators \cite{Millis} first suggested that the theoretical description of this transition must include the electron-phonon Jahn-Teller (JT) coupling, in addition to the double-exchange (DE) \cite{DE} spin-spin interaction, thus invoking polaronic degrees of freedom. Experimental signatures of small-polaron hopping have emerged from transport measurements above the FM critical temperature $T_C$ \cite{smallpol}, leading to a theoretical description of magnetoelastic polarons in terms of a small cluster of aligned Mn spins surrounding a local JT lattice distortion \cite{Millis,Roder}. 

Despite these important insights, an experimental and theoretical description of the FM transition encompassing {\em all three\/} important degrees of freedom (lattice, charge, and spin) remains incomplete. The persistence of local lattice distortions below $T_C$ is clear from neutron pair-distribution function (PDF) and x-ray absorption fine structure (XAFS) measurements \cite{pdf,xafs}. A two-fluid model for the charge degrees of freedom is consistent with the evolution of localized charges into itinerant carriers near $T_C$ \cite{Jaime}. A comparable description of the behavior of the spin system near $T_C$ is lacking, however.

In this Letter we present new muon spin relaxation ($\mu$SR) and neutron spin echo (NSE) measurements in FM La$_{1-x}$Ca$_{x}$MnO$_3$, $x \simeq 0.3$. Neutron scattering has been used to study the low-temperature magnetic properties in FM perovskites \cite{dispersion}, and has revealed a broad peak centered around zero energy transfer which coexists with Mn spin waves near $T_C$ \cite{Lynn}. An unambiguous explanation for this peak is still lacking, however. The NSE technique \cite{Mezei1} allows a direct measure of the spin-spin correlation function $S(q,t)$ at much higher resolution and for longer correlation times than are achievable with conventional neutron scattering. The muon is a local probe (bound within about 1~\AA\ of an oxygen atom in oxide materials \cite{muonpos}) and is, therefore, sensitive to spatial inhomogeneity in spin fluctuation rates. Our $\mu$SR and NSE results are consistent with a spatially distributed FM transition involving at least two interacting regions with very different spin dynamics. We are able to follow the evolution of these different regions with $\mu$SR over the temperature range $0.7~T_C \lesssim T \lesssim T_C$. 

The $\mu$SR data were taken at the Paul Scherrer Institute in Villigen, Switzerland, and at TRIUMF in Vancouver, Canada, between 10 and 300 K\@. The NSE data were taken at 275 K with the IN11 spectrometer at the Institute Laue Langevin, using incident neutron wavelengths of 4~\AA\ and 6~\AA. All measurements were performed in zero applied field ($\lesssim 1$~Oe) on polycrystalline samples. (NSE measurements below $T_C$ were impossible in zero field because the neutrons were completely depolarized by randomly oriented FM domains.) The $\mu$SR sample [$x = 0.33$, $T_C$ = $262(3)$ K from magnetization measurements] was from the same batch as that used in Ref. \cite{Heffner}. For the NSE sample $x = 0.30$ and $T_C$ = $250(3)$ K\@. In general polycrystalline materials of (La,Ca)MnO$_3$ are more compositionally homogeneous than comparable volumes of single crystals, because of relatively less Ca evaporation and fluctuating growth conditions.

The $\mu$SR data could be fit to a relaxation function $G_z(t)= G_{\rm osc}(t) + G_{\rm rlx}(t)$,  corresponding to oscillating and relaxing terms, respectively.  $G_{\rm osc}(t)$ is given by $A_{\rm osc}\exp(-t/T_2) \cos(2\pi\nu_\mu t + \phi_\mu)$, where $\nu_\mu$ is the muon precession frequency, $1/T_2$ is the inhomogeneous damping rate, and $2\pi\nu_\mu = \gamma_\mu B$, with $\gamma_\mu$ the muon gyromagnetic ratio and $B$ the local magnetic field. $G_{\rm rlx}(t)$ was first fit to the stretched-exponential form $A_{\rm rlx} \exp[-(t/T_1)^K]$, where $1/T_1$ is a characteristic spin-lattice relaxation rate and a value of $K < 1$ implies a distribution of rates. For our experiment $A_{\rm osc} + A_{\rm rlx} \approx 0.2$ independent of temperature.

The current $\mu$SR data are of greater statistical precision than previously reported \cite{Heffner}, and are taken at smaller temperature intervals near $T_C$, allowing a more refined interpretation of the relaxation phenomena. Fit parameters are shown in Figure~1.  The observed zero-field frequencies $\nu_{\mu}(T)$ tend to zero at a temperature of $266 \pm 2$ K, in agreement with $T_C$ from the magnetization data mentioned above. The amplitudes A$_{\rm osc}$ (not shown) indicate that the entire sample volume experiences growth of the sublattice magnetization. The higher statistical accuracy allows observation of the rapid decline of $K$ to about $0.2$ just below $T_C$, whereas previously it was necessary to freeze $K$ at a somewhat arbitrary value of 0.5 below about 270 K \cite{Heffner}. Because $K$ and $1/T_1$ are highly correlated, the $1/T_1$ values reported here are different from those reported earlier. 

For rapid fluctuations the local muon relaxation rate~$\lambda$ is given by
$\lambda \propto \gamma^2_{\mu} \sum_q |\delta B(q)|^2 \tau(q)$,
where $|\delta B(q)|$ is the amplitude of the fluctuating local field and $\tau(q)$ is the Mn-ion correlation time. A distribution of $\lambda$ implies that $\delta B$ and/or $\tau$ are distributed but does not determine the distributions separately, whereas $S(q,t)$ obtained from NSE measurements gives a direct measure of the distribution of $\tau(q)$ only. Figure~2 displays $S(q,t)$ at $T = 275$~K in La$_{0.70}$Ca$_{0.30}$MnO$_3$. The data can be fit to $S(q,t) = \exp(-(t/\tau(q))^\beta)$, $S(q,0) = 1$; the parameters from these fits are given in Table I\@. These results, together with the $\mu$SR data, directly confirm that the Mn-ion correlation times are spatially distributed. Note that $\beta \simeq 1$ for $q > 0.10$ \AA$^{-1}$ and $\beta < 1$ for $q \leq 0.10$ \AA$^{-1}$.

The use of a stretched exponential to describe the $\mu$SR and NSE relaxation functions is not unique and may not be physically appropriate, as we now discuss. A stretched exponential relaxation function implies a continuous distribution~$P(\lambda)$ of rates $\lambda$ defined by $\int d\lambda P(\lambda) \exp(-\lambda t) = \exp[-(\Lambda t)^{\alpha}]$, which is broad for small $\alpha$. Such a distribution is applicable to dilute spin glasses, where a distribution of energy barriers yields a broad distribution of relaxation rates, and a given spin may relax via many possible channels \cite{spinglass}.  Thus NSE measurements in dilute spin glass \underline{Cu}Mn \cite{Mezei} show no appreciable $q$ dependence of $S(q,t)$, an indication of the many relaxation channels available to a given spin. This is in marked contrast to La$_{0.70}$Ca$_{0.30}$MnO$_3$ (Fig.~2), where the $q$-dependence of $S(q,t)$ is likely due to a spatial distribution of relaxation rates, as discussed below.  We have therefore chosen to adopt a bimodal distribution~$P(\lambda) \propto A_f\delta(\lambda - \lambda_f) + A_s\delta(\lambda - \lambda_s)$, corresponding to a two-exponential fit for $S(q,t)$ and $G_{\rm rlx}(t)$.  Further justification for this is given below. 

The results from fitting $S(q,t) = A_F(q) \exp(-t/\tau_F(q)) + A_S(q) \exp(-t/\tau_S(q))$, where $A_F + A_S = 1$, are given in Table I.  The two components are labeled `fast' (small $\tau$) and `slow' (large $\tau$). The slow component sets in only for $q \leq 0.10$~ \AA$^{-1}$, and the correlation times $\tau_F$ and $\tau_S$ differ by about an order of magnitude at the smallest $q$ values. The $\mu$SR data are also well described using a two-exponential form $G_{\rm rlx}(t) = A_f \exp(-\lambda_f t) + A_s \exp(-\lambda_s t)$. Here $A_f + A_s = A_{\rm rlx}$. We have labeled the rates $\lambda_f$ and $\lambda_s$, in analogy to the NSE analysis. Again, the two exponential rates differ by at least a factor of ten, allowing good convergence of the fits. We note that non-uniqueness of the fitting function for both the NSE and $\mu$SR data is a general property of sub-exponentially decaying curves \cite{exponfit}. The justification for which function to use must therefore ultimately reside in a physical interpretation of the data. 

Figure~3 shows the temperature dependence of $A_f$, $A_s$, $\lambda_f$, and $\lambda_s$. For temperatures below 185 K the overall relaxation rate is too small to resolve two exponentials clearly, and above 273 K the exponent~$K$ approaches 1. Note that there is consistency between the NSE and $\mu$SR measurements, taking into account that  $\mu$SR  rates involve a sum over all $q$,  which in ferromagnets is most heavily weighted for $q \simeq 0$. The amplitudes of the fast and slow components near $T_C$ are roughly equal. Although the $\mu$SR data do not give precise values for $\tau_f$ and $\tau_s$, because the $\delta B(q)$ are not known, the fact that $\lambda_s/\lambda_f$ and $\tau_S/\tau_F$ (at small $q$) are both $\gtrsim 10$ is quite consistent. 

The $\mu$SR data in Fig.~3 exhibit the following important trends. The peak value of $\lambda_f$ coincides with $\nu_\mu(T) \rightarrow 0$ at $T \simeq T_C$, unlike the stretched-exponential fits in Fig.~1, where the peak values of $1/T_1$ occur at distinctly lower temperatures ($\simeq 255$ K) than where $\nu_\mu(T) \to 0$. Ordinarily the spin-lattice relaxation rate is expected to peak at $T_C$, where the susceptibility and the spin-spin correlation times reach their maxima. This does not occur for the stretched-exponential fits because the rapidly changing values of the exponent~$K$ near $T_C$ result in a changing functional form of $G_{\rm rlx}(T)$; then the derived values of $1/T_1$ are not consistent from one temperature to the next. Thus the two-exponential model function seems more appropriate. By contrast, the temperature dependence of $\lambda_s$ shows no significant maximum near $T_C$, but rises slowly below the critical temperature. Near $T_C$, $A_f$ is about 60\% of the total relaxing amplitude and gradually increases as the temperature is reduced, reaching about 75\% at $T/T_C \simeq 0.7$. The dotted lines (Fig.~3, top) are guides to the eye, and indicate that the trend of $A_f$ is to approach 100\% of the fluctuating amplitude at $T = 0$~K.

The $\mu$SR and NSE data suggest the following qualitative interpretation: there are two spatially separated regions in the sample, characterized in our measurements by very different Mn spin dynamics and temperature-dependent volumes. The muon relaxation samples these regions locally through its short-ranged dipolar coupling (matrix element $\propto r^{-6}$), predominantly to its two nearest-neighbor Mn spins. The $\mu$SR rates involve sums over all $q$, but are dominated by FM fluctuations. NSE is sensitive to the spatial configuration of the spin fluctuations through the $q$ dependence of the correlation times. The single-exponential relaxation observed for larger $q$ reflects nearly single-ion dynamics which are not associated with the FM transition, whereas the two-component relaxation evident for smaller $q$ shows that the FM dynamics are inhomogeneous. The length scale associated with this crossover is a few lattice spacings, which is roughly consistent with the `droplets' found in small-angle neutron scattering in lightly-doped manganites \cite{Hennion}. We have also studied $\mu$SR relaxation in undoped LaMnO$_3$ and CaMnO$_3$, both of which exhibit temperature-independent single-exponential relaxation above 140 K with rates~0.05--$0.10~\mu {\rm s}^{-1}$. This is very different than the behavior of either component in La$_{0.67}$Ca$_{0.33}$MnO$_3$, which is strong evidence against large-scale separation of Mn$^{3+}$ and Mn$^{4+}$ ions.

We postulate that the two observed relaxation components constitute the spin signature of magnetoelastic polarons as the system becomes FM. The more rapidly relaxing component, which shows a peak in the relaxation rate at $T_C$ (as expected for a FM phase transition), is attributed to spins inside the overlapping polaron regions. The volume fraction of these regions grows below $T_C$ as the polarons continue to grow. The second component is characterized by much slower Mn-ion relaxation rates, shows no sign of critical slowing down, and occupies a diminishing volume fraction as the temperature is lowered. We associate this component with the spins at the boundaries of, or between, the polarons. 

In this picture the interior regions are characterized below $T_C$ by relatively low (metallic) resistivity, an average Mn valence between $3^+$ and $4^+$ (corresponding to the rapid motion of charge carriers via the DE interaction), and a smaller local JT lattice distortion than is found in LaMnO$_3$. The rapid band-like carrier motion produces more rapid Mn spin fluctuations than in the less-metallic exterior regions, where spin and charge motion are frustrated by more extreme local lattice distortions. These larger distortions limit the polaron size at a given temperature. Note that the two regions must be in close proximity to one another to explain the $q$ dependence of $S(q,t)$, and thus their spins should interact. This interaction  presumably causes weak FM alignment in the less-metallic regions, and may also explain the broad peak in $\lambda_f$ around $T_C$, which is uncharacteristic of conventional critical dynamics. This breadth may also be caused by a distribution of polaron sizes (and local $T_C$ values).

We compare these data to other measurements. Our picture is conceptually similar to the two-fluid model of resistivity \cite{Jaime}, with one exception. The $\mu$SR data indicate that slowly fluctuating spins in relatively less metallic regions persist far below $T_C$, whereas in the two-fluid model the growth of the conducting charge fraction reaches essentially 100\% just below $T_C$. This can be explained by realizing that when a conducting path is reached (at $T_C$) the more resistive regions are short-circuited, even though they may still occupy a considerable volume fraction. Persistence of inhomogeneity below $T_C$ is reinforced by local probes of the lattice structure \cite{pdf,xafs,billinge}, which are consistent with the {\em gradual} loss of structural inhomogeneity below $T_C$, similar to that reflected in the spin-lattice relaxation rates reported here. Finally, it is likely that the central peak observed in inelastic neutron scattering \cite{Lynn} corresponds to our slowly relaxing component. 

We have presented evidence for two time scales in the spin system of (La,Ca)MnO$_3$, and associated these with a fine-scale spatially-distributed FM transition due to the formation and accretion of magnetoelastic polarons. Realistic theories of these slow inhomogeneous spin dynamics have yet to be developed. The spatial distribution of spin fluctuation rates is probably related to the disordered spatial distribution of La and Ca ions and the corresponding local fluctuations of the local lattice distortions, which influence the spin subsystem via magnetoelastic coupling. A theoretical model which includes disorder and couples JT pseudo-spin variables with the Mn-spin degrees of freedom may shed light on the mystery of the slow, inhomogeneous spin dynamics reported here. 

Work at Los Alamos was performed under the auspices of the U.S. DOE\@. Work elsewhere was supported in part by the U.S. NSF, Grant no.~DMR-9731361 (U.C. Riverside), and by the Dutch Foundations FOM and NWO (Leiden).

\begin{table}[tbh]
\caption{Fitting parameters for NSE spin-spin correlation function $S(q,t)$ in La$_{0.70}$Ca$_{0.30}$MnO$_3$, $T = 275$ K\@. Symbols are defined in the text.}
\label{Table 1}
\begin{tabular}{ccccccc}
$q$ (\AA$^{-1}$) & $\beta$ & $\tau$ (ps) & $A_F$ & $A_S$ & $\tau_F$ (ps) & $\tau_S$ (ps) \\ \hline
0.04 & 0.59(4) & 10.6(5) & 0.63(5) & 0.37(5) & 4.2(6) & 41(8)\\
0.05 & 0.60(4) & 8.1(3) & 0.70(4) & 0.30(4) & 4.0(4) & 40(4) \\
0.07 & 0.78(6) & 5.7(2) & 0.67(9) & 0.33(9) & 3.5(5) & 15(3) \\
0.10 & 0.82(4) & 3.6(1) & 0.55(21) & 0.47(22) & 1.8(6) & 7(2) \\
0.14 & 1.06(7)& 2.1(1) & -- & -- & -- & -- \\
0.20 & 1.03(9)& 1.22(1) & -- & -- & -- & -- 
\end{tabular}
\end{table}

\break \begin{figure} 
\caption{Temperature dependence of the muon precession frequency $\nu_\mu$ (top), average spin lattice relaxation rate $1/T_1$ (middle) and exponent $K$ (bottom) for the stretched exponential relaxation function $G_{\rm rlx}$.}
\end{figure}

\begin{figure}
\caption{Spin-spin correlation function $S(q,t)$ measured with NSE spectroscopy. The data for $q \geq 0.10$ \AA$^{-1}$ were taken at incident wavelengths of 4~\AA\ and 6~\AA\ and hence extend to shorter times than data for $q < 0.10$ \AA$^{-1}$, taken only at 6~\AA.}
\end{figure}

\begin{figure}
\caption{Temperature dependence of $\mu$SR amplitudes $A_f/A_{\rm rlx}$ and $A_s/A_{\rm rlx}$ (top) and their respective relaxation rates $\lambda_f$ (middle) and $\lambda_s$ (bottom) for a two-exponential fit. The subscripts $f$ (filled symbols) and $s$ (open symbols) refer to fast and slow Mn relaxation rates, respectively (see text). The dotted lines in the top frame show the trends extrapolated to $T = 0$ K\@. Note the different temperature scales for the top and lower two frames.}
\end{figure}

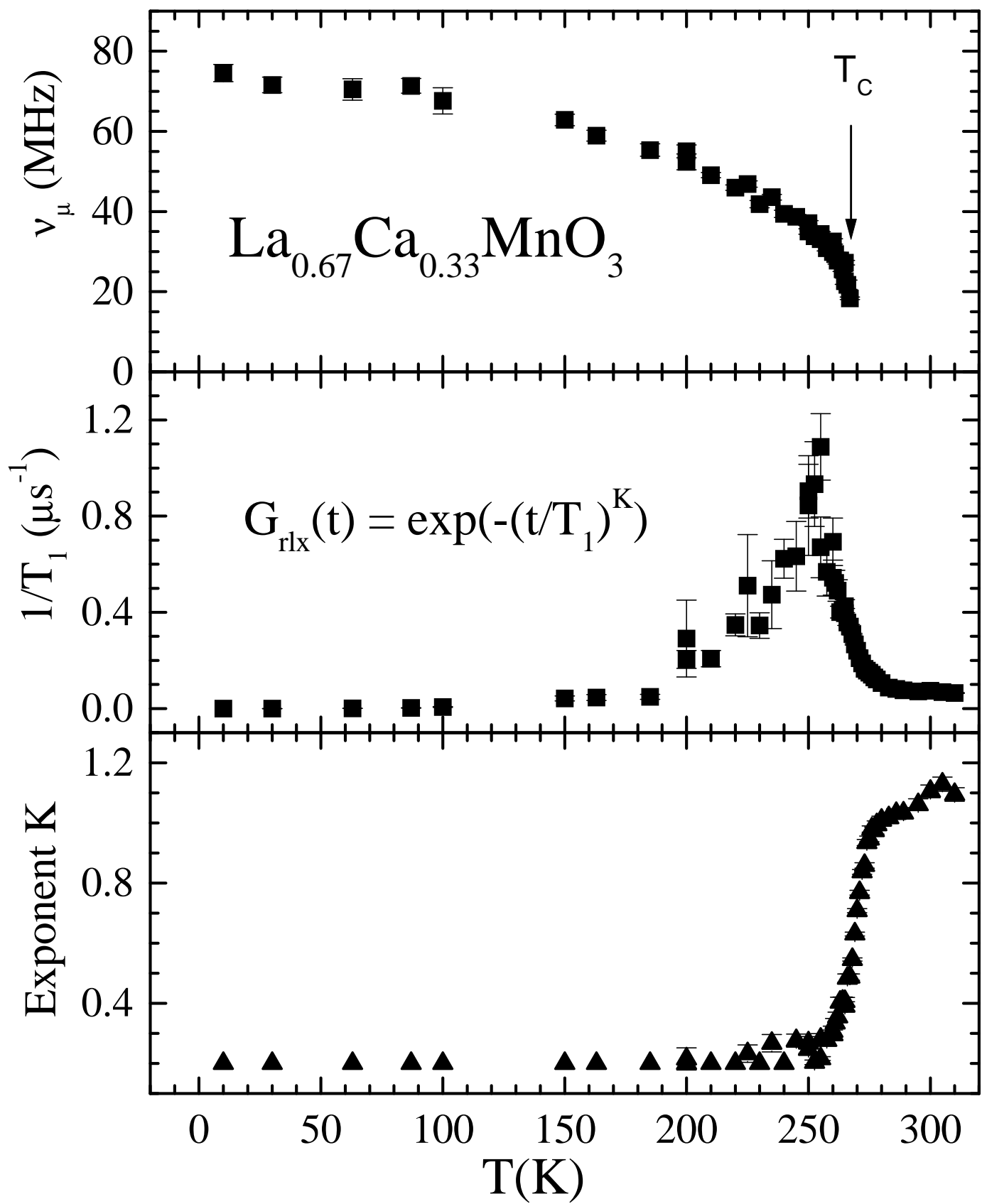
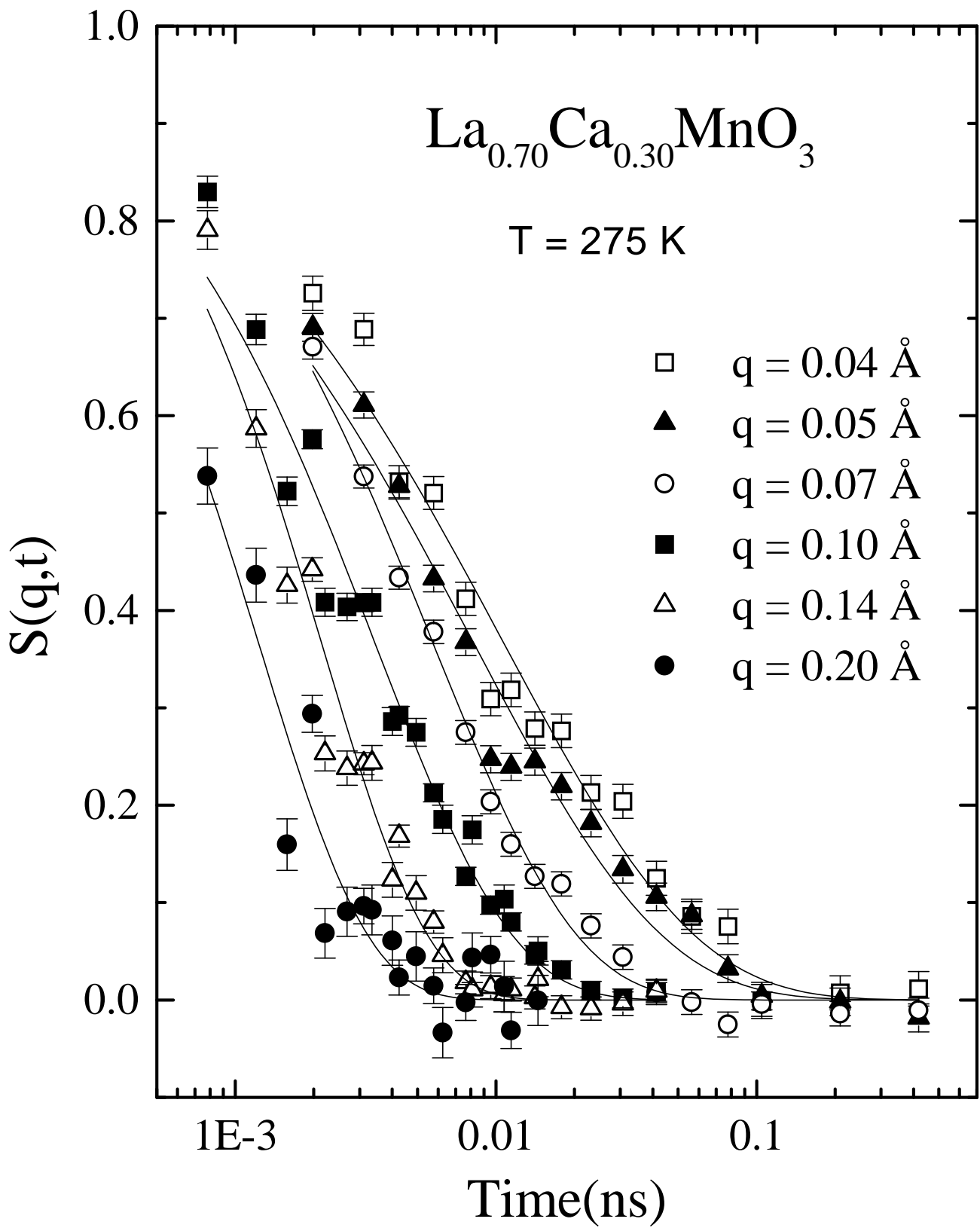
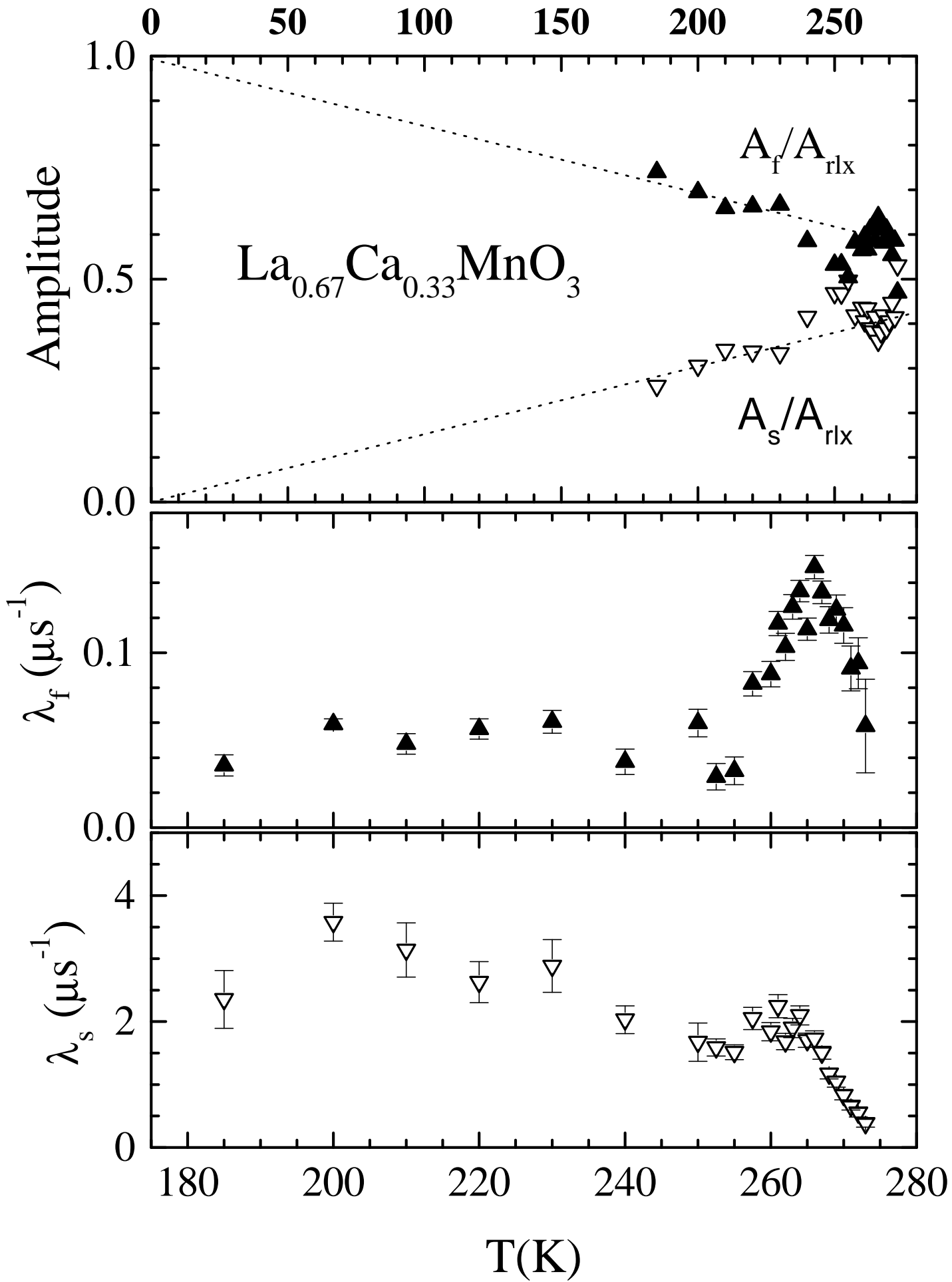
\end{document}